\documentstyle[12pt,epsfig,espcrc2c]{article}

\newcommand{\be}{\begin{eqnarray}}
\newcommand{\ee}{\end{eqnarray}}

\title{QCD at Finite Density and Color Superconductivity }

\author{E.V.Shuryak\address{ Department of Physics and Astronomy, SUNY Stony Brook, NY 17794}
        \thanks{Supported in part by the US Department
 of Energy under Grant No. DE-FG02-88ER40388.               }
}
       
\begin{document}

\begin{abstract}
\end{abstract}
\maketitle

\section{Brief history}
 ``Prehistoric'' 
 QCD-based calculations dealed with plasma-type phenomena like
  Debye screening etc \cite{Shu_80}, both for finite T and density.
 Early ideas about  Color
Superconductivity 
 \cite{earlysuper} were based on simple observation:
 unlike electrons, quarks of 
{\it different} {colors} are attracted to each other even
  by Coulomb forces. Due to Cooper
  instability
any  small attraction is enough: however the
superconducting
gap was only
 $\Delta\sim 1 \ MeV$, and applicability of perturbative QCD
was in doubt.

 My interest was initiated by finding 
\cite{SSV_diquarks} that in the instanton liquid model 
even without $any$ quark matter,
the  {\em ud scalar diquarks} are very
deeply bound,  by amount comparable to
 the constituent quark mass. So, 
phenomenological manifestations \cite{diquarks}
 of such diquarks have in fact  deep dynamical roots: they follow
 from the same basic dynamics
as the  ``superconductivity'' of the QCD vacuum, the chiral
 ($\chi$-)symmetry breaking. These
spin-isospin-zero
   diquarks are related to pions, and should be quite 
  robust element of nucleon (octet baryons) structure\footnote{As opposed to $\Delta$
  (decuplet) baryons.} . 

Another argument for deeply bound diquarks comes from 
bi-color ($N_c=2$) theory: in it the scalar diquark is degenerate
with pions.  By
 continuity from $N_c=2$ to $3$, 
 a trace of it
 should exist in real QCD\footnote{
Instanton-induced interaction strength in diquark channel is
$1/(N_c-1)$ of that for $\bar q\gamma_5 q$ one. It is the same at
$N_c=2$, zero for large  $N_c$, and is   exactly in
 between for $N_c=3$.}.
  
Explicit calculations with
instanton-induced forces for  $N_f=2,N_c=3$ QCD
have been made in two simultaneous\footnote{Submitted to hep-ph on the
  same day.} papers
 \cite{RSSV,ARW}. Indeed, a  
  very robust
Cooper pairs and
  gaps
$\Delta\sim$ 100 MeV were found. From then on, the field is booming.

 This phase
(called CSC2) has the same symmetries as
discussed before \cite{earlysuper}: the $\chi$-symmetry is restored but
color group is broken by the diquark condensate, acting like Higgs VEV
 of the Standard Model.
New variety of color superconductor, CSC3 with {\em Color-flavor locking}
 exists for 3 or more light flavors
 $N_f=3$, see
sect.5.
At asymptotically high densities
the
perturbation theory must become right, see
 section 6. 
Finally we will have discussion of some outstanding issues in sect.7. 

\section{Physics Overview}
 The QCD phase diagram, as we
 understand it today, is shown in fig.1 at the
  $\mu-T$ plane  (baryonic chemical potential -temperature)
At small $T$,$\mu$ there is
ordinary hadronic matter with broken chiral
symmetry. 
The point M (from ``multi-fragmentation'')
is the endpoint of the nuclear liquid-gas phase transition.
 At the (hypothetical) critical point E
the first order line either continue as 2-nd order 
 (for $m_u,m_d=0$) or disappears (for finite  masses): according to
recent proposal \cite{SRS} it can be found in real heavy ion
collisions. QDQ (quark-diquark) phase is  hypothetical \cite{RSSV2}: I
have no time to speak about it here.
The main point is locations of the  two superconducting phases, 
CSC2 and CSC3. At T=0 going to
large $\mu$, the $\chi$-symmetry seem to be first recovered in CSC2,  and then
broken  again in CSC3.
\begin{figure}[h]
\begin{center}
\vskip -1.cm
\epsfig{file=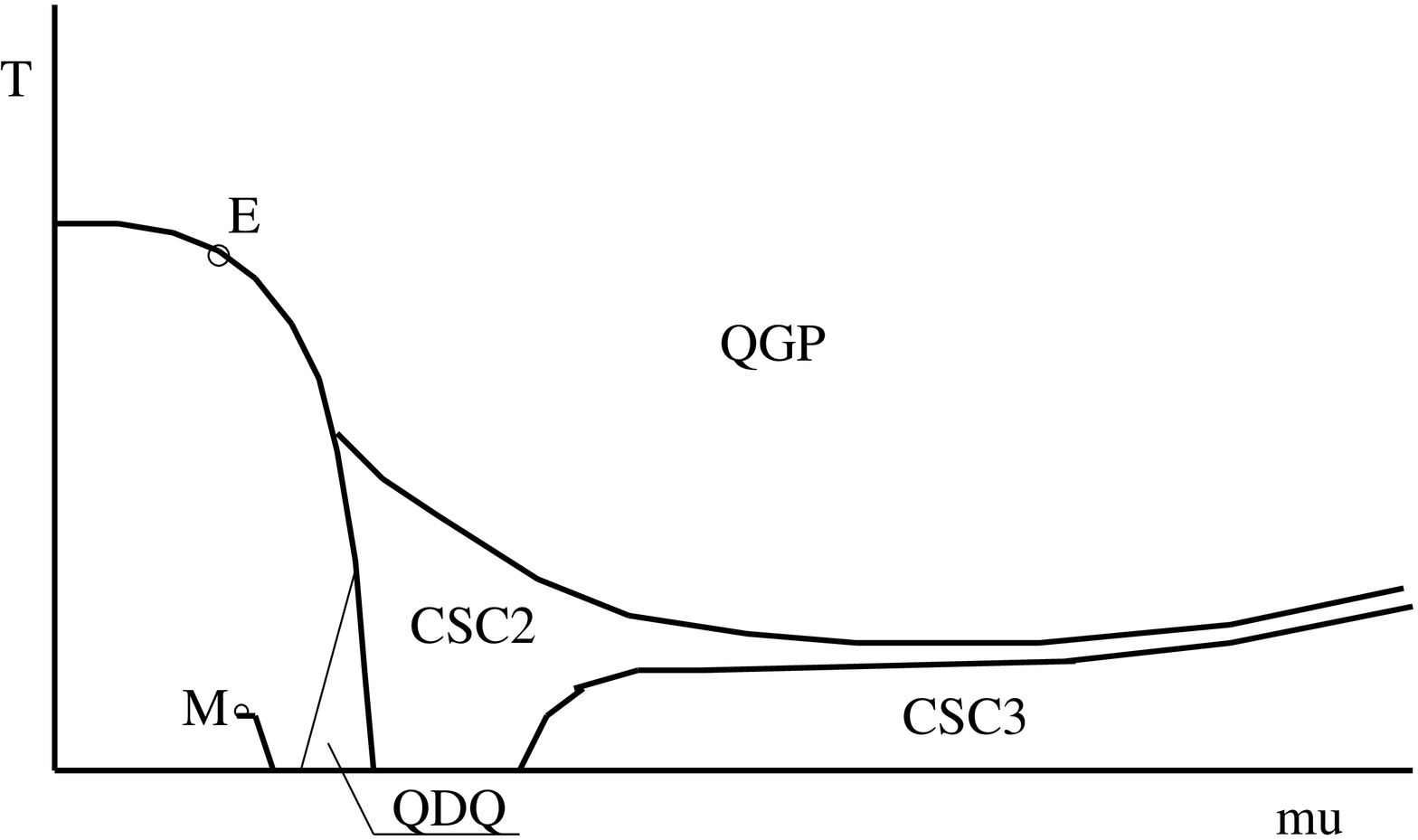,width=4cm,angle=270}
\end{center}
\label{fig_phases}\vskip -1.cm
\end{figure}
\noindent

 Let me then explain few major
 physics points. 
Why is there a transition from {\em particle-hole} 
 to  {\em particle-particle} pairing?  The following
 figure
(dispersion
curves $\omega(k)$) for quarks in vacuum and superconductor explains it: it
is better to have a gap at
 the surface of the {Fermi} sphere rather than the {Dirac} sea.

\begin{figure}[h]

\leavevmode
\epsfxsize=2.in
 \begin{minipage}[c]{3.in}
 \centering 
\includegraphics[width=1.5in, angle=270]{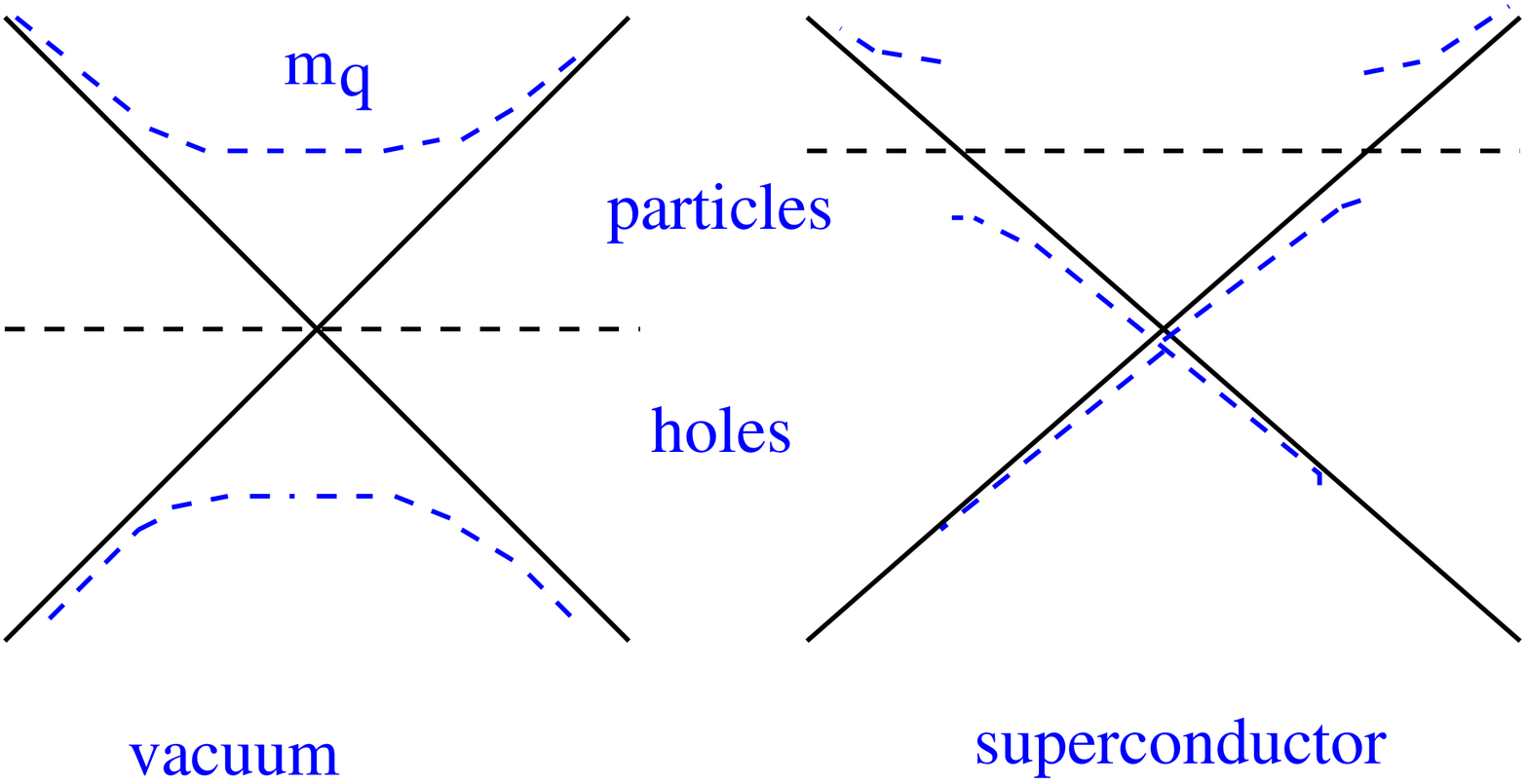}
 \end{minipage}
\vskip -1.cm
\end{figure}

 {\em Why instantons?}
The reasons are: (i) They are  
  the {\em strongest} non-perturbative effect known; (ii) Unlike OGE,
  they $do$
 explain quantitatively  $\chi$-symmetry breaking in
    vacuum; (iii) 
  Anomaly cannot be  eliminated  by finite density, so
  tunneling
leads to level crossing  at 
 the surface of the {Fermi} sphere as well.

%

Instantons create  the following amusing {\em  triality}: 
there are three attractive channels 
 which compete: (i)
the
{instanton-induced} attraction in
$\bar q q $ channel
leading to  $\chi$-symmetry breaking.
(ii) the
{instanton-induced} attraction in
$ q q $ which leads to color superconductivity.
(iii)
the
{\em light-quark-induced} attraction of $\bar I I $, which
leads to pairing of instantons
into {``molecules''} and a Quark-Gluon Plasma (QGP) phase without 
$any$ condensates.

 {\em How the calculations are actually done?}. 
 Analytically, mostly in 
  the mean field approximation,  similar to the original BCS
  theory
in Gorkov formulation. 
Total thermodynamical potential consists of 
{\em ``kinetic energy''} of the quark Fermi gas, including 
 mass operators of two types (shown in figure below).
The  {\em ``potential energy''} in such approximation is the
interaction Lagrangian convoluted  with all possible condensates.
 For example, instanton-induced one with $N_f=3$ leads to  two types of
diagrams shown in Fig.4, with (a)
{$<\bar q q>^3$} and (b)  {$< q q>^2 <\bar q q>$}.
Then one minimizes the potential over all condensates and get {\em gap equations}:
algebra may be involved because  masses/condensates are
{\em  color-flavor matrices.} 
\begin{figure}[h]
\leavevmode
\epsfxsize=5.cm
\vskip -0.2in
 \begin{minipage}[c]{2.in}
 \centering 
\includegraphics[width=.8in, angle=270]{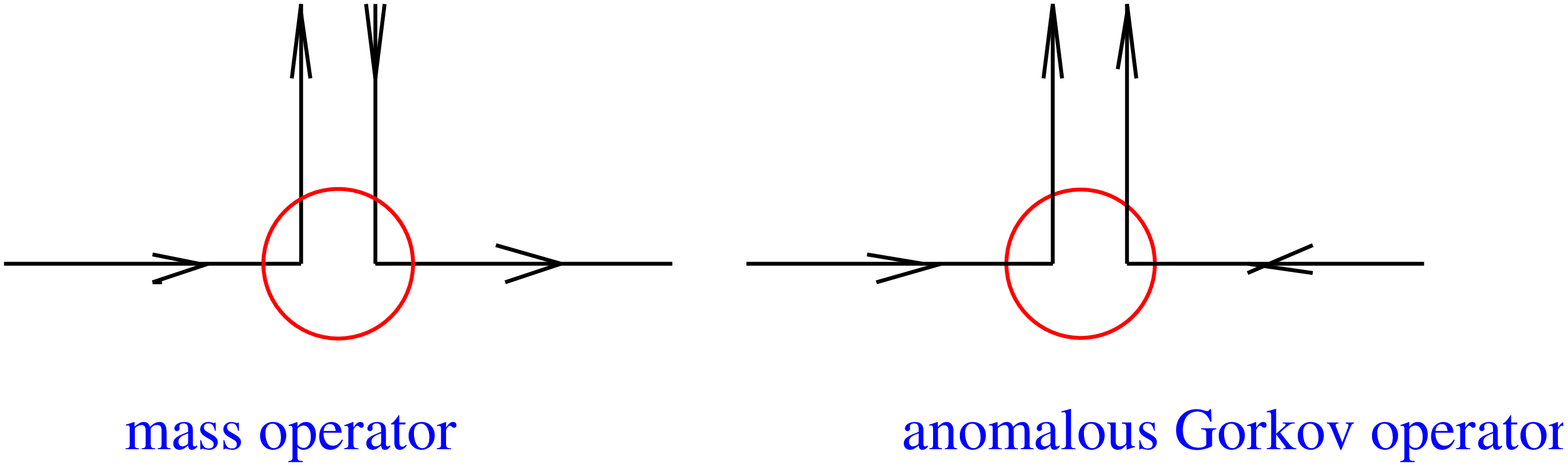}
\centering
\includegraphics[width=1.in, angle=270]{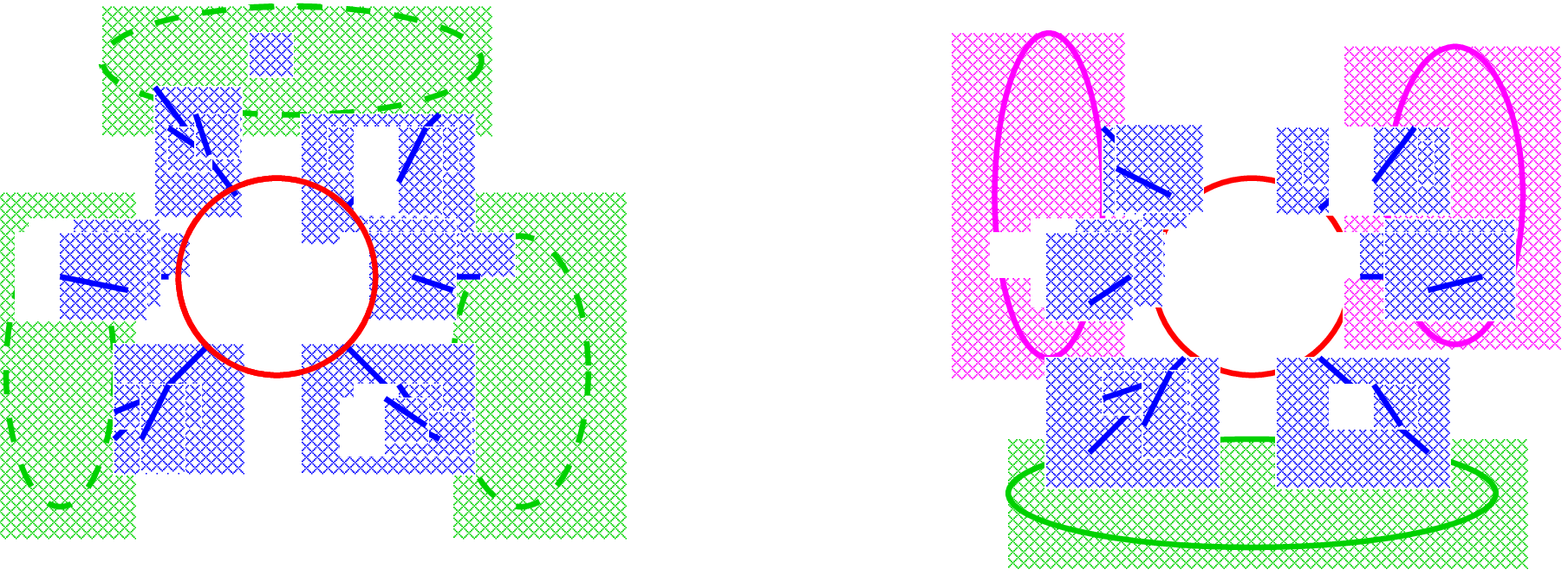}
\vskip -0.5in
\end{minipage}
\end{figure}

%

\section{ {Two colors: a very special theory  }}
One reason it is special is well known to lattice community: its
fermionic determinant is $real$ even for non-zero $\mu$,
 which makes simulations possible.
Early works by Karsch, Dagotto et al (of mid-80's!) made sense, but were
looked at only now. 

 The major interest to this theory
is related the so called {\em Pauli-Gursey symmetry}, due to which
diquarks are
{ degenerate} with mesons.
The $\chi$-symmetry breaking is 
$SU(2N_f)\rightarrow Sp(2N_f)$, 
 for $N_f=2$ the coset 
$ K=SU(4)/Sp(4)=SO(6)/SO(5) =S^5$.
 Those 5 massless modes are
 pions plus scalar diquark S and its anti-particle $\bar S$. The corresponding
sigma model was worked out in \cite{RSSV}: for further development 
see \cite{KST}.  As argued in  \cite{RSSV}, in this theory
the
critical value of transition to Color Superconductivity is simply
$\mu=m_\pi/2$,  the  diquark condensate is 
  just rotated $<\bar q q>$ one, and the gap is the constituent quark mass.
Recent lattice
works   \cite{2col} and
instanton liquid simulation \cite{S_dens} display it in great details, building
confidence for other cases.

 
\section{Two flavor QCD: the CSC2 phase}
\begin{figure}[h]
\vskip -1.cm
\centering 
\includegraphics[width=4.in]{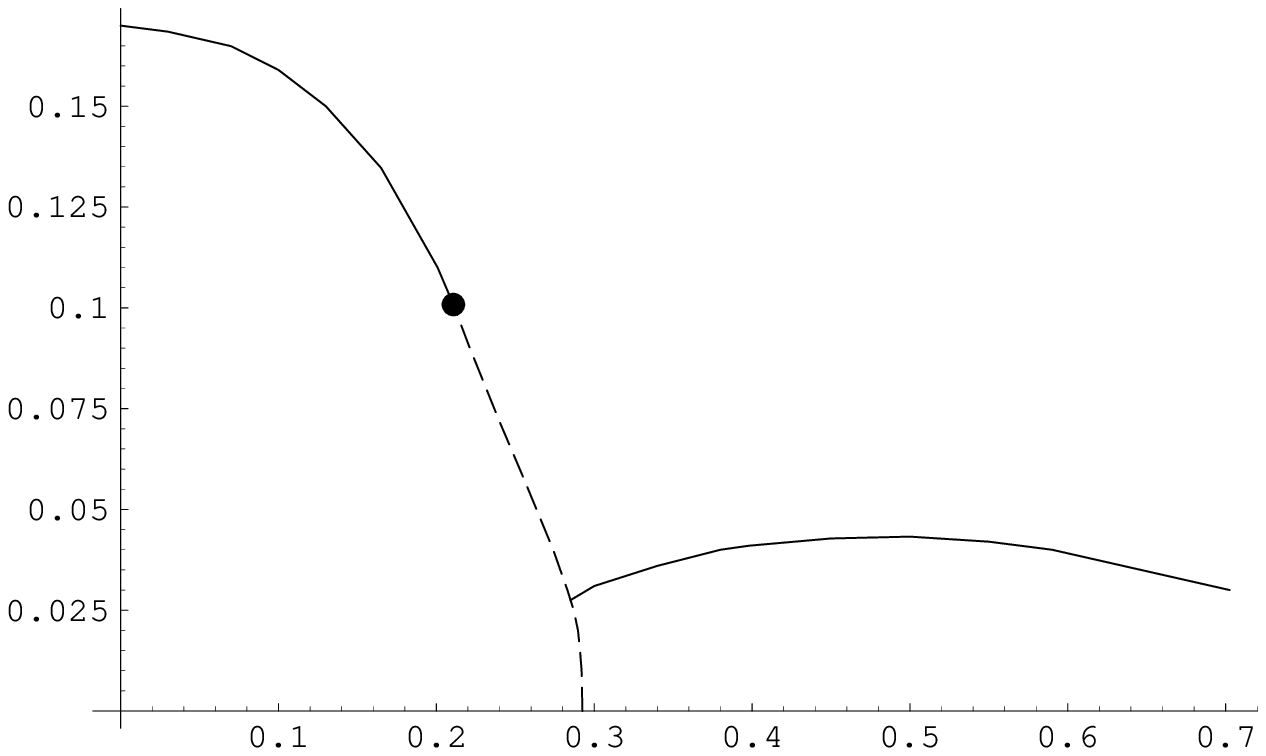}
\vskip -8.cm
\end{figure}
This phase diagram \cite{BR} is
a rare example of calculated T-$\mu$ one: The 1st
order line is dashed, and the 2nd order ones are solid lines.
 Most
studies of this theory  \cite{RSSV,ARW,CD} are at T=0. 
In all these works one more  possible 
phase (intermediate between vacuum and CSC2), {\it Fermi gas of
  constituent quarks}, with
{both $M,\Delta\neq 0$} - was unstable. However in last more refined
calculation \cite{RSSV2} it obtains a small window, as shown
by the dashed line on the following figure. Its features are amusingly close
to those of nuclear matter: but it isn't, of course: to get nucleons
one should go outside the mean field. First attempted to do so in
\cite{RSSV2} was for  another cluster - the $\bar I I$
molecules. At T=0 it is however only 10\% correction to previous
results, but is dominant as T grows.


\begin{figure}[h]
\vskip -.3cm
     \begin{minipage}[c]{3.in}
     \centering 
\includegraphics[width=2.6in]{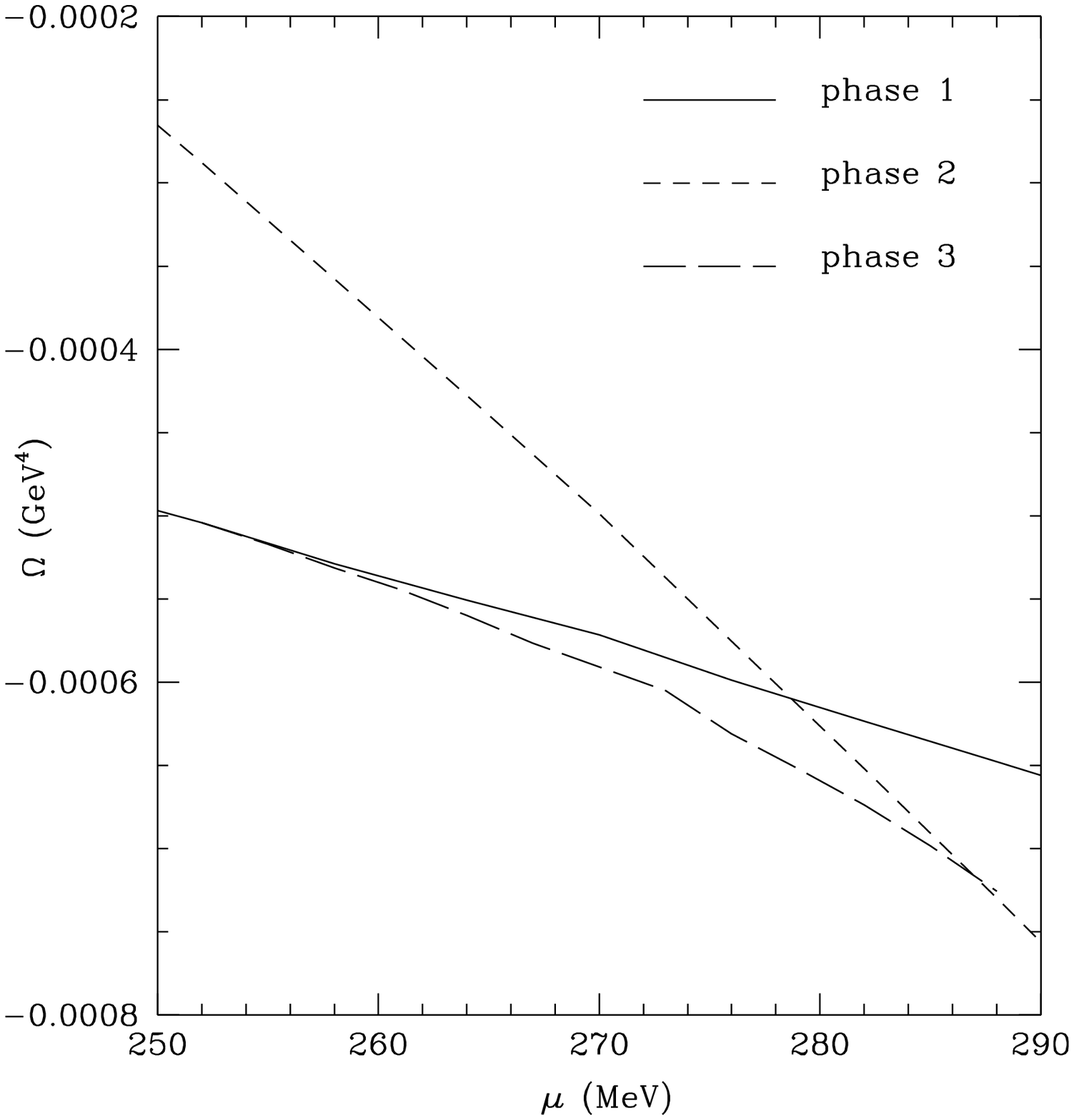}
\vskip -.6cm
      \end{minipage}
\end{figure}

 
\section{$N_f=3$ QCD: the CSC3 phase}

The {\em color-flavor locking} \cite{ARW2} means that
diquark condensate has the  structure
$\langle q_i^a C q_j^b\rangle = \bar\Delta_1 \delta_{ia}\delta_{bj}
 + \bar\Delta_2 \delta_{ib}\delta_{ja}$, where ij are color and ab
 flavor indices. It is very symmetric,  reducing
$SU(3)_c SU(3)_f \rightarrow  SU(3)_{diagonal}$.
 It was verified in \cite{ARW2} 
    for the OGE interaction, and for
instanton-induced one in
\cite{RSSV2}: probably it is  always true for that theory.
Gaps $\delta_i$ and masses $\sigma_i$, following from instanton-based
calculation \cite{RSSV2}, are shown as a function of $\mu$ in the
following figure 
\begin{figure}[h]
\vskip -1.cm
  \centering 
\includegraphics[width=2.6in]{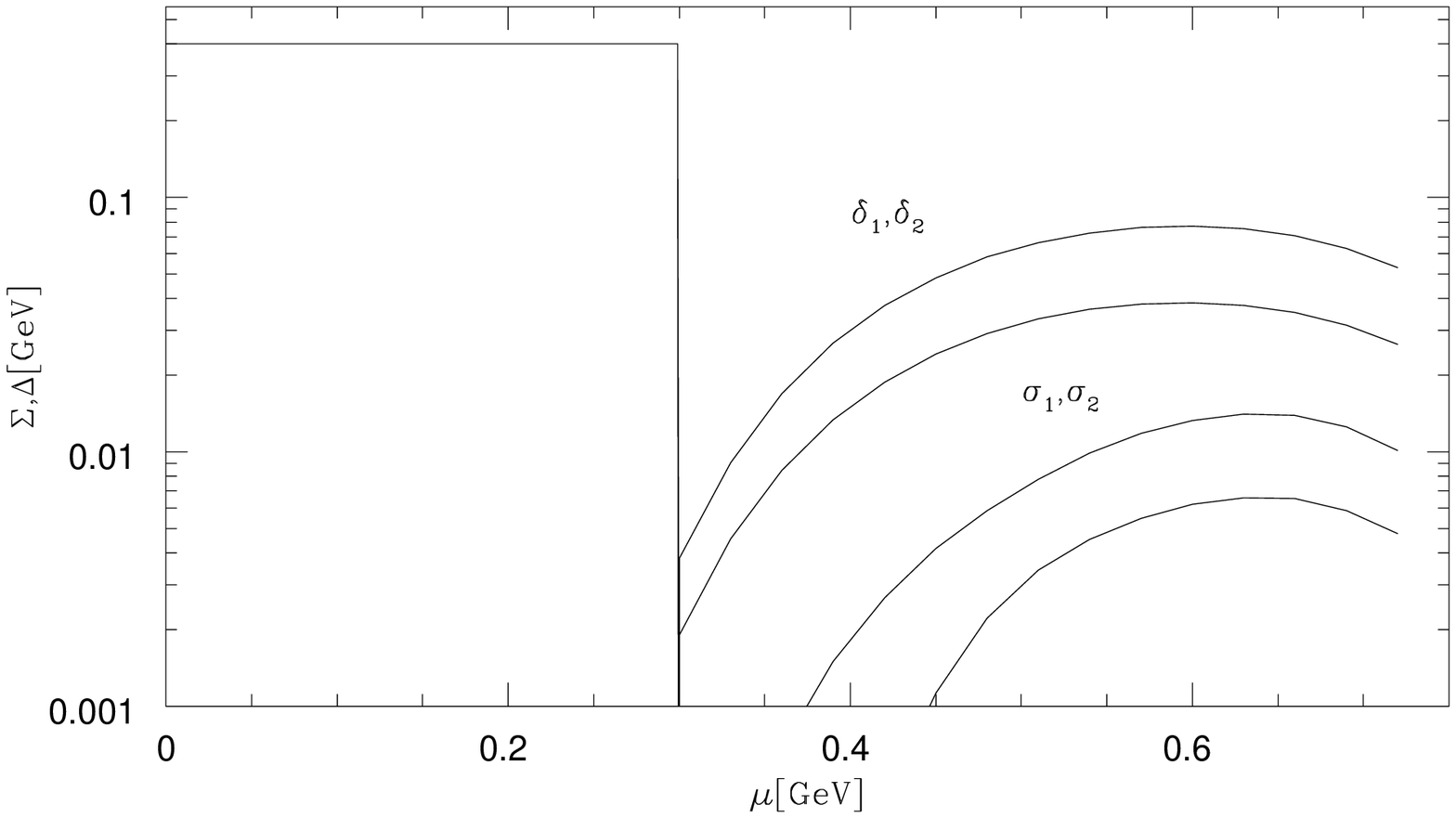}
\vskip -1.cm
\end{figure}

{\bf {Two} plus {strange} flavor QCD ($m_s\neq 0$)}
was studied in several papers \cite{RSSV2,2+1}.  Just
kinematically,  us,ds Cooper pairs with zero
momentum
is difficult to make: for $\mu_{u,d}=\mu_{s}$ the momenta
$p^F_{u,d}\neq p^F_{s}$.
Instantons generate also dynamical operator $m_s (\bar u\bar d) (ud)$. 
Resulting behavior is as shown in our first figure.
 
\section{ Asymptotically large  densities }
   At high densities $\mu> 1 GeV$ instantons are Debye-screened \cite{Shu_82}, as
   well as electric (Coulomb) OGE. So
  {\em magnetic} gluons overtake electric ones   \cite{son}.
 $Magnetically$ bound Cooper pair is
 interesting by itself, as a rare example: 
one has  to take care of {\em time delay effects} with
 Eliashberg eqn, etc. Angular integral leads to
second log in the gap equation, leading to   unusual answer:
$ \Delta \sim \mu \ exp( -3\pi^2 /\sqrt{2}g)$
which implies that
 the gap {\em grows} indefinitely with $\mu$ \footnote{ 
Numerical details for all densities can be found in recent work
\cite{SW_high}.} and pQCD becomes finally justified. However, it is
the case for
huge densities, with $\mu >10 GeV$ or so.


\section{Physics issues under discussion}

{\bf The hadron-quark continuity}. As 
 pointed out in \cite{SW_cont},  the CSC3 phase  
not only has the same  symmetries as hadronic 
matter (e.g. broken $\chi$-symmetry), but also very
similar
excitations. 
8 gluons become 8 {\em massive} vector mesons, 
3*3 quarks become 8+1  ``baryons''. The
8 massless pions  remain massless\footnote{Very exotic 3d objects,
  ``super-qualitons'' \cite{HRZ}, the skyrmions made of 
pions are among the excitations.}. Furthermore, photon and gluons
are combined into a {\em massless } $\gamma_{inside}$.
Can these phases be $distiquished$, and should there be $any$ phase transition
(in $N_f=3$ theory)?
 {\em Is it a superconductor}, after all?

I think the answers still is ``yes''.
 For example, if one puts a  piece of CSC3
into a magnet, it may levitate: although
$\gamma_{inside}$
is massless,  the magnet uses $\gamma_{outside}$ field and a part of it
is expelled\footnote{The same would happen with
  a small piece of   Weinberg/Salam vacuum, if one can make
magnet with ``original''
(``outside'') field.}. 

Let me finish with few homework questions.
What is the role of confinement in all these transitions?
What is  nuclear matter for different quark masses, anyway?
 Do we have other phases in between, like diquark-quark phase
  or (analog of) K condensation, or different crystal-like phases?
 Is there indeed a (remnant of) the tricritical point which we can
find experimentally? And, above all, 
 {\em How to do finite density calculations on the lattice?}


\begin{thebibliography}{20}
\bibitem{Shu_80} E.V.Shuryak, Phys.Rept.61,71(1980)
\bibitem{earlysuper} S. C. Frautschi (Erice78),F. Barrois, 
Nucl. Phys. B129, 390 (1977),D. Bailin and A. Love, Phys. Rep. 107,
325 (1984)
\bibitem{SSV_diquarks}T.Schaefer, E.V.Shuryak., J.Verbaarschot
Nucl. Phys. B412, 143 (1994)
\bibitem{diquarks} 
 M.~Anselmino et al., Rev. Mod. Phys. 65, 1199 (1993).
\bibitem{RSSV} R. Rapp, T. Sch\"afer, E. V. Shuryak
    and M. Velkovsky
Phys. Rev. Lett. {\bf 81} (1998) 53.
\bibitem{ARW} M. Alford, K. Rajagopal and
F. Wilczek 
Phys. Lett. {\bf B422} 247 (1998).
\bibitem{ARW2} M. Alford, K. Rajagopal and
F. Wilczek, 
hep-ph/9804403.
\bibitem{SRS} M. Stephanov, K. Rajagopal, E.V.Shuryak,
 Phys.Rev.Lett.81( 1998), hep-ph/9806219
\bibitem{Shu_82}
E.V.Shuryak, Phys.Lett. 79B,135 (1978), 
Nucl. Phys. B203,140 (1982)
 \bibitem{RSSV2} R. Rapp, T. Sch\"afer, E. V. Shuryak
    and M. Velkovsky,hep-ph 9904353, Ann.of Phys., in press.
 \bibitem{BR}J.Berges and K.Rajagopal, Nucl.Phys. B538, 215 (1999) 
 hep-ph/9804233 
\bibitem{CD}
G. W. Carter, D.I.Diakonov,
hep-ph/9812445.
\bibitem{son} D.T. Son, Phys.Rev. D59:(1999); hep-ph/9812287
\bibitem{SW_high}T.Schafer and F.Wilczek  hep-ph/9906512
\bibitem{2col}S. Hands, J.B.Kogut, M.-P.Lombardo,
S.E. Morrison, hep-lat/9902034; M.-P.Lombardo
 hep-lat/9907025; 
\bibitem{KST} J.B. Kogut, M.A. Stephanov, D. Toublan, hep-ph/9906346
\bibitem{S_dens} T.~Sch{\"a}fer, Phys. Rev {\bf D57} (1998) 3950.
\bibitem{2+1} T. Sch\"afer and F. Wilczek,
hep-ph/9903503.M. Alford, J. Berges, and K. Rajagopal,
hep-ph/9903502.
\bibitem{SW_cont}T.Schafer and F.Wilczek, Phys.Rev.Lett. 82,3956 1999) 
 hep-ph/9811473 
\bibitem{HRZ}Deog Ki Hong, M.Rho and
I.Zahed,   hep-ph/9906551
\end{thebibliography}
\end{document}